\documentclass[twoside,11pt]{article}

\usepackage{jmlr2e}
\usepackage{amsmath,amssymb,natbib}
\usepackage{bm}
\usepackage{graphicx}
\usepackage{color}
\usepackage{float}
\usepackage{units}
\usepackage{url}
\usepackage{verbatim}
\usepackage{geometry}
\usepackage{slashbox}
\usepackage{calligra}
\usepackage{epstopdf}
\usepackage{booktabs,caption,fixltx2e}
\usepackage[flushleft]{threeparttable}
\usepackage{multirow}
\usepackage{kbordermatrix}
\usepackage{tabularx}

\newcommand{\iid}{\stackrel{\mathrm{iid}}{\sim}}
\newcommand{\ind}{\stackrel{\mathrm{ind}}{\sim}}

\newcommand\independent{\protect\mathpalette{\protect\independenT}{\perp}}
\def\independenT#1#2{\mathrel{\rlap{$#1#2$}\mkern2mu{#1#2}}}

\ShortHeadings{Bayesian Nonparametric Functional Mixture Estimation}{Savitsky}

\begin{document}

\title{Bayesian Nonparametric Functional Mixture Estimation for Time-Series Data, With Application to Estimation of State Employment Totals}

\author{\name Terrance D. Savitsky \email savitsky.terrance@bls.gov \\
       \addr U. S. Bureau of Labor Statistics\\
       Office of Survey Methods Research\\
       Washington, DC 20212, USA}

\maketitle

\begin{abstract}
The U.S. Bureau of Labor Statistics use monthly, by-state employment totals from the Current Population Survey (CPS) as a key input to develop employment estimates for counties within the states.  The monthly CPS by-state totals, however, express high levels of volatility that compromise the accuracy of resulting estimates composed for the counties.  Typically-employed models for small area estimation produce de-noised, state-level employment estimates by borrowing information over the survey months, but assume independence among the collection of by-state time series, which is typically violated due to similarities in their underlying economies.  We construct Gaussian process and Gaussian Markov random field alternative functional prior specifications, each in a mixture of multivariate Gaussian distributions with a Dirichlet process (DP) mixing measure over the parameters of their covariance or precision matrices.  Our DP mixture of functions models allow the data to simultaneously estimate a dependence among the months and between states.  A feature of our models is that those functions assigned to the same cluster are drawn from a distribution with the same covariance parameters, so that they are similar, but don't have to be identical.  We compare the performances of our two alternatives on synthetic data and apply them to recover de-noised, by-state CPS employment totals for data from $2000-2013$.
\end{abstract}


\section{Introduction} \label{motivation}
The Current Population Survey (CPS), a household survey administered by the Census Bureau for the Bureau of Labor Statistics (BLS), publishes time-indexed, state-level estimates of variables relating to employment, such as the employment total.  The time-indexed employment estimates express a high degree of volatility, so that BLS will apply independently state-specified regression models to reduce the volatility before proportioning these estimates to labor market area or county as part of their local area unemployment survey (LAUS) program. This approach assumes independence among the state-indexed collection of time series.  State administrators frequently express concern that the resulting county-level estimates composed from the state-level regression models are overly volatile, in that they believe much of the month-over-month movement is noise, not signal that they would be able to attribute to underlying economic drivers.  So the LAUS program seeks a more accurate means to extract de-noised, state-level estimates of employment totals such that state administrators are readily able to ascribe economic drivers to the resulting county-level employment trends.

Bayesian nonparametric alternatives, such as the Gaussian process construction implemented by \citet{VanRiiHarJylVeh14}, treat the collection of time series as sums of latent functions and a noise process.  The structure of the model is focused to the latent functions, which are generated from a multivariate Gaussian distribution under covariance constructions that regulate the smoothness properties of the resulting functions \citep{rasm:2006}. \citet{VanRiiHarJylVeh14} specify their Gaussian process prior as independent over the states. Dependence among a set of noisy functions over the states using a Gaussian process prior formulation would require vectorizing the functions into a single function or vector of length equal to the number of states multiplied by the number of time-indexed measurement waves.  This approach would be computationally prohibitive were one to sample the full joint posterior distribution as computation time scales with order equal of the cube of the resulting vector length.  The rendering of sets of functions into a single vector is also not a natural way to perform inference because the ordering of the states may notably impact both global and local smoothness properties of the resulting vector and this approach assumes a continuous transition between state functions.

We offer an approach that builds on the independent nonparametric Bayesian functional models by probabilistically tying the set of latent functions together under a marginal nonparametric mixture prior, where the mixing measure receives a Dirichlet process (DP) prior \citep{ferguson:1973}.
Our modeling framework is similar to \citet{Gelf:Kott:Mac:baye:2005}, but their approach imposes the Dirichlet process prior directly on the functions, so that co-clustered functions must be exactly the same, which we find to be too restrictive for working with our CPS employment data application.  We will, instead, impose the DP prior on the covariance matrix parameters of the generating Gaussian distribution, such that state functions which are co-clustered are drawn from a Gaussian distribution with the same parameters, but are not required to be exactly equal.

We formulate our nonparametric mixture approach using two alternative prior constructions to parametrize the covariance matrix of a multivariate Gaussian prior imposed on the set of latent functions; 1. A Gaussian process prior; 2. An intrinsic Gaussian Markov random field prior.  These alternative formulations are developed and their properties contrasted in Section~\ref{models}.  Schemes to sample the posterior distributions under each of our two functional prior alternatives are sketched in Section~\ref{computation}, where we comment on their computational properties.  We compare their performance, both in fitting the functions and uncovering a clustering structure used to generate the functions in Section~\ref{simulation}.  Our nonparametric mixture models are applied for estimation of state-level employment totals obtained from the CPS in Section~\ref{application}, followed by a concluding discussion in Section~\ref{discussion}.

\section{Models for Functional Data} \label{models}
\subsection{CPS Employment Count Data}
Our data are composed for of $T  = 158$ months of direct estimates of employment totals for each of $N = 51$ states (including the District of Columbia) published by the BLS in the Current Population Survey. Our modeling interest is focused on reducing estimation volatility to improve the precision of the monthly, state-level employment totals.  The employment totals are influenced by the underlying economic conditions of the states.  We, therefore, expect a correlation of trends in the mix of industrial, service and agricultural economic activities among states, on the one hand, with those expressed in their employment total time series, on the other hand.  State policy makers seek context around their reported employment and unemployment levels in order to understand the drivers for the estimated trends.  We will later demonstrate that performing inference on clusters of distributions generating the set of state-indexed functions provides such context.

We denote the set of $(T=158)\times 1$ vectors of survey direct estimates for a collection of $N = 51$ states with $\{\mathbf{y}_{i}\}_{i=1,\ldots,N}$.  We would like to estimate de-noised, smooth functions, $\{\mathbf{f}_{i}\}$, from the $\{\mathbf{y}_{i}\}$.  We estimate the dependence among the collection of state employment totals through a prior construction that permits a grouping or clustering of covariance (under a Gaussian process (GP) prior specification) or precision parameters (under an intrinsic Gaussian Markov random field (iGMRF) prior specification), that we index by state.  These state-indexed covariance or precision parameters are used to generate the latent $\{\mathbf{f}_{i}\}$.

We first outline GP and iGMRF formulations with simpler, global covariance and precision parameters, respectively, $\left(\bm{\theta}, \kappa\right)$, not indexed by state, to illustrate the properties of functions generated under these prior specifications.  Employing a single set of covariance or precision parameters assumes the functions are exchangeably drawn. 
We will subsequently extend both of these models using nonparametric mixture formulations over the states for $\{\bm{\theta}_{i}\}$ and $\{\kappa_{i}\}$ under the GP and iGMRF models, respectively, to allow for dependence among the functions.

\subsection{Gaussian process Model} \label{gpmod}
\begin{figure}[!ht]
\begin{center}
\includegraphics[width=3.5in,height=3.0in]{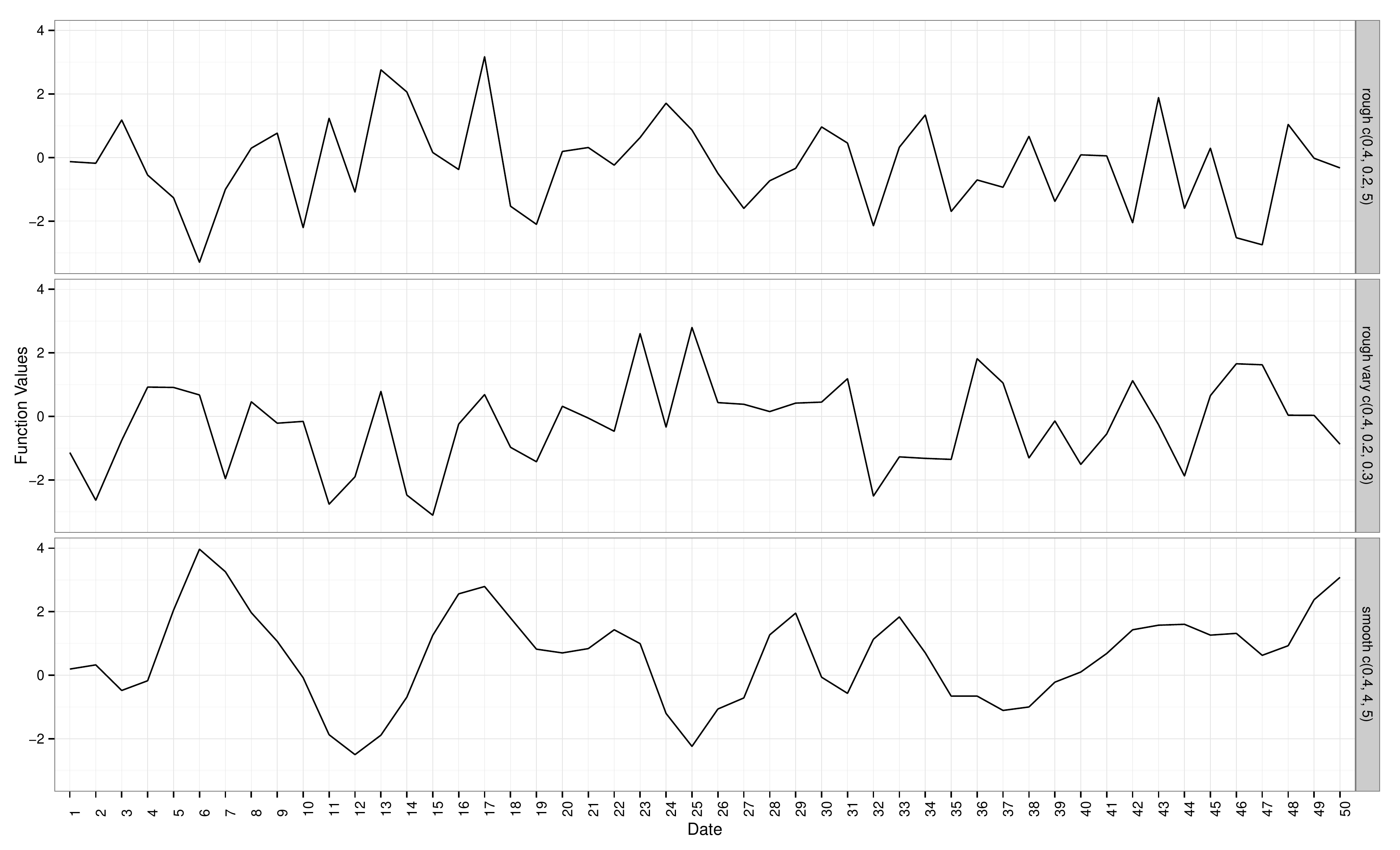}
\caption{Randomly-drawn functions, $\mathbf{f}$, from a GP using a rational quadratic covariance function under distinct values for $\bm{\theta} = \left(\theta_{1},\theta_{2},\theta_{3}\right)$.  The top panel employs $\bm{\theta} = \left(0.4,0.2,5.0\right)$ to generate a ``rough" surface, while the bottom
panel specifies $\left(0.4,4.0,5.0\right)$ to characterize a ``smooth" surface,
and the middle panel employs $\left(0.4,0.2,5.0\right)$ to produce a rough surface characterized by a high-degree of length-scale variation.}
\label{gpdraws}
\end{center}
\end{figure}

Our Gaussian process (GP) model is parameterized through the covariance matrix, $\mathbf{C}\left(\bm{\theta}\right)$, where $\bm{\theta}$ are estimated by the data and control the trend, length scale (or wavelength) and variation in generated functions, $\{\mathbf{f}_{i}\}$.  We specify a GP formulation in the following probability model:
\begin{subequations}
\label{globalgp}
\begin{align}
\mathop{\mathbf{y}_{i}}^{T\times 1} &\ind \mathcal{N}_{T}\left(\mathbf{f}_{i},\tau_{\epsilon}^{-1}\mathbb{I}_{T}\right),~i=1,\ldots,N \label{gplike}\\
\mathbf{f}_{i} &\iid \mathcal{N}_{T}\left(\mathbf{0},\mathop{\mathbf{C}\left(\bm{\theta}\right)}^{T\times T}\right)\label{fprior}\\
\theta_{1},\ldots,\theta_{P}|a,b &\iid \mathcal{G}a\left(a = 1.0,b = 1.0\right)\\
\tau_{\epsilon}|c,d &\iid \mathcal{G}a\left(c = 1.0 ,d = 1.0\right),
\end{align}
\end{subequations}
where $\bm{\theta} = \left(\theta_{1},\ldots,\theta_{P}\right)$. The GP model specifies a formula for the covariance matrix, $\mathbf{C}\left(\bm{\theta}\right)$, with,
\begin{eqnarray*}
\mathbf{C}\left(\bm{\theta}\right) &=& \left(C_{f_{j},f_{\ell}}\right)_{j,\ell \in\left(1,...,T\right)}\\
C_{f_{j},f_{\ell}} &=& \frac{1}{\theta_{1}}\left(1 + \frac{\left(t_{j}-t_{\ell}\right)^{2}}{\theta_{2}\theta_{3}}\right)^{-\theta_{3}},
\end{eqnarray*}
for $P = 3$.  The particular covariance formula we use is known as the rational quadratic, which may be derived as a scale mixture of more commonly-used squared exponential kernels, $1/\theta_{1}\exp\left((t_{j}-t_{\ell})^{2}/\theta\right)$, with the inverse of the length scale parameter, $\theta^{-1}$, which controls function periodicity, distributed under a Beta distribution with hyperparameters $\left(\theta_{3},\theta_{2}^{-1}\right)$ \citep{rasm:2006}.  The vertical magnitude of a function drawn under the rational quadratic covariance formula is directly controlled by $\theta_{1}$, while $\theta_{2}$ controls the mean length scale or period of the function, and $\theta_{3}$ controls smooth deviations from $\theta_{2}$.  As $\theta_{3}\uparrow\infty$, this formulation converges to a single squared exponential kernel with length-scale, $\theta_{2}$.  Figure~\ref{gpdraws} displays a randomly-generated function under each of $3$ distinct value settings for for $\bm{\theta} = \left(\theta_{1},\theta_{2},\theta_{3}\right)$ of the rational quadratic covariance formula to provide insight into how $\bm{\theta}$ produces distinct features in $\mathbf{f}$.  Our choice of the rational quadratic covariance formulation is intended as a parsimonious specification for parameterizing the use of a single covariance matrix that is able to discover multiple length-scales or wavelengths in a function, in lieu of using a set of squared exponential covariance terms (e.g. in an additive formulation) with each term specialized to a particular length scale.

The rational quadratic formula also produces smooth surfaces, $\{\mathbf{f}_{i}\}$, because they are constrained to be differentiable at all orders.  We prefer covariance formulas that render smooth functions, $\{\mathbf{f}_{i}\}$,  because the smoothness property more identifiably separates signal from rough, non-differentiable, noise.  A primary goal for this work is to extract de-noised functions from noisy time series. The GP of Equation~\ref{globalgp} under the rational quadratic covariance formula is very flexible, allowing the estimation of functions, $\{\mathbf{f}_{i}\}$, to be non-linear of any order within the space of smooth functions.  This GP is equivalent to an infinite dimension polynomial spline, even though the parameterization of the GP is through the covariance matrix, rather than the mean.  A linear mean model may be re-parameterized as a zero mean GP by marginalizing over the regression coefficients, which demonstrates that linear surfaces are a subset of the space of smooth functions parameterized in Equation~\ref{globalgp} \citep{neal:2000}.
The structure of Equation~\ref{globalgp} is contained in $\mathbf{f}_{i}$, so that the $\{\mathbf{y}_{i}\}$ are assumed to conditionally-independent, given $\{\mathbf{f}_{i}\}$.
Standard errors for the monthly CPS direct estimates were not available to us.  We treat the model noise precision, $\tau_{\epsilon}$, as unknown, such that it is estimated by the data, because this parameter influences the amount of signal, represented by $\mathbf{f}$, extracted from the noisy response, $\mathbf{y}$, and is a feature of the GP model.  Our probability model in Equation~\ref{globalgp}, that specifies a Gamma prior for $\tau_{\epsilon}$, is fully identified.

\subsection{Intrinsic Gaussian Markov Random Field Model} \label{gmrfmod}
An iGMRF prior may be viewed as a probabilistic local smoother composed from differences in function values, which in our case, are indexed by time.  A typical iGMRF specification is placed on the first difference approximation to the first derivative, $\Delta f_{ij} = f_{i,j+1} - f_{i,j} \iid \mathcal{N}\left(0,\kappa^{-1}\right)$, where $\kappa$ is the earlier discussed precision parameter that determines the (vertical) scale of variation among the first differences.  This approach uses nearest neighbors defined from first differences, $\left(f_{i,j-1},f_{i,j+1}\right)$, to encode the time-indexed dependence structure among the ${f}_{ij},~j = 1,\ldots,T$ for each domain, $i \in (1,\ldots,N)$.  Using first differences may produce an excessively rough (though continuous) surface that over-fits the data (by making it difficult to separately model signal and noise processes), so we employ a prior construction based on the second difference approximation to the first derivative, $\Delta^{2}f_{ij} = \Delta\left(\Delta f_{ij}\right) \iid \mathcal{N}\left(0,\kappa^{-1}\right)$.  This prior on second differences produces the joint distribution, $\mathbf{f}_{i} \iid \kappa^{\frac{T-2}{2}}\exp\left(-\frac{\kappa}{2}\mathop{\sum}_{j = 1}^{T-2}\left(f_{ij} - 2f_{i,j+1} + f_{i,j+2}^{2}\right)\right) = \kappa^{\frac{T-2}{2}}\exp\left(-\frac{1}{2}\mathbf{f}_{i}^{'}\mathbf{R}\mathbf{f}_{i}\right)$, where $\mathbf{R} = \kappa\mathbf{Q}$ specifies a band-diagonal precision matrix with non-zero entries, $\left(Q_{j,j-2}, Q_{j,j-1}, Q_{j,j}, Q_{j,j+1}, Q_{j,j+2}\right) = \left(1,-4,6,-4,1\right)$ for $2 < j < T - 2$ for non-boundary parameters \citep{rue:held:2005}. Under this construction, $\mathbf{R}$ is rank-deficient (of rank $T-2$) as the rows sum to $0$ since it is composed from second differences, so that the joint distribution is improper; in particular, the prior for the $T \times 1,~\mathbf{f}_{i}$, is invariant to the addition of any second order polynomial because the prior supplies no information about such polynomials.  We may view the joint distribution as the product of a proper distribution on the space of $T - 2$ differences (by employing the Moore-Penrose pseudo inverse, $\mathbf{R}^{-}$, and $\vert\mathbf{R}\vert$ as the product of the $T-2$ non-zero eigenvalues of $\mathbf{Q}$) and an improper, noninformative prior on the order $2$ polynomials. These Gaussian Markov random field priors specified through a precision matrix have the property that $f_{ij} \independent f_{ik} \vert \mathbf{f}_{i,-jk} \leftrightarrow R_{jk} = 0$, which allows for a parsimonious $\mathbf{Q}$, from which we specify a \emph{proper} set of full conditional distributions that we use for posterior sampling,
\begin{subequations}
\label{globalgmrf}
\begin{align}
f_{ij}\vert \mathbf{f}_{i,-j},\kappa &\sim \mathcal{N}\left(-\frac{1}{Q_{jj}}\mathop{\sum}_{k:k\sim j}Q_{jk}f_{ik},\left(\kappa Q_{jj}\right)^{-1}\right) \\
&= \mathcal{N}\left(\frac{4}{6}\left(f_{i,j+1} + f_{i,j-1}\right) - \frac{1}{6}\left(f_{i,j+2} + f_{i,j-2}\right), \left(6\kappa\right)^{-1}\right),
\end{align}
\end{subequations}
where $\{k:k\sim j\}$ denotes the set of neighboring time points, $\{k\}$, of time, $j$.  The prior mean for each $f_{ij}$ is composed as a weighted average of its order $2$ nearest neighbors.  \citet{rue:held:2005} refer to this construction as a random walk prior of order $2$ or $RW2(\kappa)$.  One may equivalently parameterize, $\mathbf{R} = \kappa\left(\mathbf{D}-\rho{\mathbf{\Omega}}\right)$, for diagonal weight matrix, $\mathbf{D}$, with (diagonal) entries equal to those of $\mathbf{Q}$.  Element, $j$, of $\mathbf{D}$, counts the number of neighbors of time, $j$, where the function value at a time point with more neighbors will have a relatively higher precision.  Adjacency matrix, $\mathbf{\Omega}$, is specified with $0's$ on the diagonal and off-diagonal elements are filled with the negative of values from the off-diagonal elements of $\mathbf{Q}$ and encode adjacency relationships among the time points \citep{Bane:Wall:Carl:frai:2003}.  Under this parameterization, $-1 < \rho < 1$, is the autocorrelation parameter that shrinks the mean to $0$, but produces a proper joint prior for $\mathbf{f}_{i}$.  The $RW2(\kappa)$ prior construction, however, sets $\rho = 1$.  As with Equation~\ref{globalgp}, we specify $\tau_{\epsilon} \sim \mathcal{G}a\left(c = 1.0, d = 1.0\right)$.

The GP construction of Equation~\ref{globalgp} includes parameters $\left(\theta_{2},\theta_{3}\right)$ for the length-scale that are estimated from the data, while the iGMRF prior hard codes the length scale in $\mathbf{Q}$, suggesting more estimation flexibility for the GP, which we assess in Section~\ref{simulation}.

\subsection{Accounting for Dependence Among Functions}
We introduce an extension of Equation~\ref{globalgp}, which indexes the GP covariance function parameters, $\{\bm{\theta}_{i}\} = \{\left(\theta_{i1},\ldots,\theta_{iP}\right)\}$, by state, $i \in \left(1,\ldots,N\right)$, to permit their probabilistic clustering with,
\begin{subequations}
\label{dpgp}
\begin{align}
\mathbf{f}_{i} &\sim \mathcal{N}_{T}\left(\mathbf{0},\mathbf{C}\left(\bm{\theta}_{i}\right)\right)\label{fdplike}\\
\bm{\theta}_{1},\ldots,\bm{\theta}_{N}|G &\iid G \\
G &\sim \mbox{DP}(\alpha, G_{0}),
\end{align}
\end{subequations}
where $\{\bm{\theta}_{i}\}_{i=1,\ldots,N}$ receive a random distribution prior, $G$, drawn from a Dirichlet process (DP), specified with a concentration parameter, $\alpha$, a precision parameter that controls the amount of variation in $G$ around prior mean, $G_{0}$. The base or mean distribution, $G_{0}$, is typically constructed as parametric; in our case, $G_{0} = \mathop{\prod}_{p = 1}^{P}\mathcal{G}a\left(1.0,1.0\right)$, which we used under the global GP model of Equation~\ref{globalgp} parameterizing a single vector, $\bm{\theta}$, for all states. Equation~\ref{dpgp} describes a mixture model of the form, $\mathbf{f}|G \iid \int \mathcal{N}_{T}\left(\mathbf{0},\mathbf{C}\left(\bm{\theta}\right)\right)
G\left(d\bm{\theta}\right)$, where $G$ is the mixing measure over the $P\times 1$ covariance parameters, $\bm{\theta}$.

We examine the clustering property of the DP by expressing it in the discrete, stick breaking form of \citet{sethuraman:1994},
\begin{equation}\label{stick}
G = \mathop{\sum}_{h=1}^{\infty} p_{h}\delta_{\bm{\theta}^{\ast}_{h}},
\end{equation}
a countably infinite mixture of weighted point masses, where ``locations", $\bm{\theta}^{\ast}_{1},\ldots,\bm{\theta}^{\ast}_{M}$, index the unique values for the $\{\bm{\theta}_{i}\}$, where $M \leq N$ (states) that we interpret as clusters.  The maximum number of clusters assigns each state to its own cluster, which countably increases with $N$. We record state cluster memberships with $\mathbf{s} = \left(s_{1},\ldots,s_{N}\right)$ where $s_{i} = \ell$ denotes $\bm{\theta}_{i} = \bm{\theta}^{\ast}_{\ell}$ so that $\left(\mathbf{s},\{\bm{\theta}^{\ast}_{m}\}\right)$ provides an equivalent parameterization to $\{\bm{\theta}_{i}\}$ and we recover $\bm{\theta}_{i} = \bm{\theta}^{\ast}_{s_{i}}$.  We conduct posterior sampling with the cluster locations and assignments, rather than directly sampling $\{\bm{\theta}_{i}\}$, as the former produces notably better mixing because it separates re-assignments to clusters from updates to the values.  The weight, $p_{h} \in (0,1)$, is composed as $p_{h} = v_{h}\mathop{\prod}_{k = 1}^{h-1}\left(1-v_{k}\right)$ where $v_{h}$ is drawn from the beta distribution, $\mathcal{B}e\left(1,\alpha\right)$.  This construction provides a prior penalty on the number of mixture components.  A higher value for $\alpha$, however, will generate smaller values for $\{v_{h}\}$, and hence, ``breaks" off more clusters (with unique locations), $\{p_{h}\}$, from the unit stick.   We place a further prior, $\alpha \sim \mathcal{G}\left(1,1\right)$, to allow posterior updating in recognition of the relatively strong influence it conveys on the number of clusters formed \citep{escobar:1995}.

The DP construction assumes exchangeability of the $\left(\bm{\theta}_{i}\right)$, \emph{a priori}, given random measure, $G$, but the almost surely discrete construction of the DP produces estimates which are not exchangeable, \emph{a posteriori}. The prior specification for cluster assignments, $\left(s_{i}\vert\mathbf{s}_{-i}\right)$, (under our re-parameterization to $\{\mathbf{s},(\bm{\theta}^{\ast}_{m})_{m=1,\ldots,M}\}$ achieved by marginalizing over $G$), induces a uniform probability for co-clustering among the states.  Let $\mathbf{C}^{\tau}\left(\bm{\theta}_{m}^{\ast}\right) = \mathbf{C}\left(\bm{\theta}_{m}^{\ast}\right) + \left(1/\tau_{\epsilon}\right)\mathbb{I}_{T}$ after integrating out $\mathbf{f}_{i}$ from Equation~\ref{gplike} that produces the marginal likelihood, $\mathbf{y}_{i}\vert \mathbf{s},\left(\bm{\theta}^{\ast}_{m}\right) \sim \mathcal{N}_{T}\left(\mathbf{0},\mathbf{C}^{\tau}\left(\bm{\theta}_{s_{i}}^{\ast}\right)\right)$.  If the likelihood values for two states, $j$ and $k$, $\mathcal{N}_{T}\left(\mathbf{y}_{j}|\mathbf{0},
\mathbf{C}^{\tau}\left(\bm{\theta}^{\ast}_{s}\right)\right)$ and $\mathcal{N}_{T}\left(\mathbf{y}_{k}|\mathbf{0},\mathbf{C}^{\tau}\left(\bm{\theta}^{\ast}_{s}\right)\right)$, are both relatively high for assignment to cluster, $s_{j} = s_{k} = s$, due to underlying similarities in their economies or due to other factors related to their closeness in their geographic locations, then the posterior probability for co-clustering states $j$ and $k$ will be relatively high (versus uniform, \emph{a priori}). This posterior estimation mechanism conducts \emph{unsupervised} (probabilistic) clustering.   Sharper (lower posterior variance) estimates over the space of clusterings may be obtained, particularly when the number of observations are higher (than our $N=51$), by indexing either the weights or locations in Equation~\ref{stick} to include predictors in lieu of our unsupervised formulation.

An analogous extension is specified from Equation~\ref{globalgmrf} with,
\begin{subequations}
\label{dpgmrf}
\begin{align}
f_{ij}\vert \mathbf{f}_{i,-j},\{\kappa_{i}\}_{i=1,\ldots,N} &\sim \mathcal{N}\left(-\frac{1}{Q_{jj}}\mathop{\sum}_{k:k\sim j}Q_{jk}f_{ik},\left(\kappa_{i} Q_{jj}\right)^{-1}\right) \\
\kappa_{1},\ldots,\kappa_{N}|G &\iid G \\
G &\sim \mbox{DP}(\alpha, G_{0}),
\end{align}
\end{subequations}
that may be expressed as $f_{ij}|\mathbf{f}_{i,-j}, G \iid \int \mathcal{N}\left(-\frac{1}{Q_{jj}}\mathop{\sum}_{k:k\sim j}Q_{jk}f_{ik},\left(\kappa Q_{jj}\right)^{-1}\right)
G\left(d\kappa\right)$, marginalizing over $\{\kappa_{i}\}_{i=1,\ldots,N}$.  We specify base distribution, $G_{0} = \mathcal{G}a\left(1,0.1\right)$, (which generates locations, $\{\kappa^{\ast}_{m}\}$) as weakly informative with a large variance.  We select these hyperparameter settings because they produce a larger prior variance for $\bm{\kappa}$ in Equation~\ref{dpgmrf} than for $\mathbf{\Theta}$ under Equation~\ref{dpgp}.  The parameterization of the precision matrix of each function using only a single parameter ($\kappa$) makes the posterior sensitive to the prior specification.  As with Equation~\ref{dpgp}, we sample $\bm{\kappa}$ indirectly through cluster assignments, $\mathbf{s}$, for the $N \times 1$ states, and location values $\kappa^{\ast}_{1},\ldots,\kappa^{\ast}_{M}$.

The stick breaking formulation illustrated in Equation~\ref{stick} to induce the unknown measure, $G$, may be easily generalized beyond the DP by varying the form for the Beta prior assigned to $v_{h}$; for example, the Poisson-Dirichlet (PD) alternative of \citet{Pitm:Yor:two-:1997} includes an additional ``discount" parameter, $\delta$, along with the DP concentration parameter, $\alpha$, to specify the hyperparameters of the Beta distribution for $v_{h}$ that tends to generate more locations, \emph{a priori}, inducing a less informative prior for the number of clusters (which is random under DP and PD constructions). Large clusters, however, receive more prior weight under the PD than for the DP.  The DP may be extracted as a special case of the PD and both may be located within a larger class of generalized gamma processes \citep{Lijo:Mena:Prns:cont:2007}. The PD and DP turn out to produce nearly identical posterior results under our data application due to a high level of information in the data, so we focused our exposition on the simpler model.

\section{Computation}\label{computation}
The mixtures of Gaussian processes formulation of Equation~\ref{dpgp} is far more computationally-intensive than than the mixtures of iGMRFs model presented in Equation~\ref{dpgmrf} because computing the cholesky decomposition of the $T \times T$ covariance matrix is $\mathcal{O}(T^{3})$.  We report computation time comparisons in Section~\ref{simulation} where we will, however, also demonstrate that the mixtures of GPs is more robust in discovering the correct number of generating clusters than is the mixtures of iGMRFs.  So we are motivated to mitigate the computational burden associated to drawing posterior samples under a GP prior on the functions.  A typically-used approach to improve computation for a GP probability model employs a sparse approximation to the GP that utilizes some subset of the time points as ``inducing" inputs \citep{VanRiiHarJylVeh14}.  We expect functions, $\{\mathbf{f}_{i}\}$, for our CPS data to \emph{each} express multiple length scales because our estimation period encompasses over one hundred months, such that a sparse approximation may fail to capture key features.  So we adapt a recently developed posterior sampling algorithm of \citet{wang:2013} to our mixtures of GPs that generates a sequence of proposals for locations, $\{\bm{\theta}^{\ast}_{m}\}$, in a lower dimensional space using a subset of the $T$ time points to build the GP covariance matrix, $\mathbf{C}$, which reduces computation time.  Yet, sample draws under this method will be from the exact posterior distribution, rather than from a sparse approximation.

Please see Appendix~\ref{AppMain} for an exposition of the posterior sampling algorithms used for both the mixtures of GPs and iGMRFs models, which we have implemented in the \verb*#growfunctions# package for \citet{R}, using C++ to speed computation.  The package is fully documented and available on CRAN for downloading and includes the data used in this paper.

\section{Simulation Study}\label{simulation}
We conduct a simulation study with the goals to: $1).$ Assess and compare the accuracy of estimating de-noised functions between our GP and iGMRF formulations; $2).$ Assess and compare the accuracy of detecting clusters or sub-groups of de-noised functions. We generate each latent synthetic function, $\mathbf{f}$ (from which we render the observed time series, $\mathbf{y}$) under different parameterizations than either of our models as a means to assess the relative adaptability and robustness of our estimation models to real-world data generating processes.

Our first simulation focuses on generating relatively complicated surfaces from a mixture of $K = 2$ length scales specific to each of $M = 3$ clusters.  We randomly (and disjointly) allocate $N = 100$ domains to $M = 3$ clusters for $T = 158$ time points, and generate de-noised functions, $\{\mathbf{f}_{i}\}$, in each cluster using the addition of two squared-exponential terms,
\begin{equation*}
\mathbf{f}_{i} \sim \mathcal{N}_{T}\left(\mathbf{0},\mathbf{C}\left(\bm{\theta}^{\ast}_{1,s_{i}}\right) +
\mathbf{C}\left(\bm{\theta}^{\ast}_{2,s_{i}}\right)\right).
\end{equation*}
The covariance
formula used to generate each cell of the covariance matrix associated to each squared exponential term is given by,
\begin{equation*}
C_{f_{ij},f_{i\ell}}\left(\bm{\theta}^{\ast}_{k,s_{i}}\right) = \frac{1}{\theta^{\ast}_{k,s_{i},1}}\exp\left(\frac{(t_{ij}-t_{i\ell})^{2}}{\theta^{\ast}_{k,s_{i},2}}\right),
\end{equation*}
where ($j,\ell$) is the cell indicator in the $T\times T$, $\mathbf{C}$. Term indicator, $k \in \{1,2\}$, denotes the covariance term with associated locations, $\{\theta^{\ast}_{k,m,1},\theta^{\ast}_{k,m,2}\}$, for a squared exponential covariance (which has two location parameters). Label $(k,m,1)$ indicates a location
parameter that controls the vertical scale for covariance term $k$ in cluster $m$, whereas $(k,m,2)$ denotes a location parameter that controls the length scale or degree of smoothness.  We recall that the vector, $\mathbf{s} = \left(s_{1},\ldots,s_{N}\right),~s_{i} \in (1,\ldots,M)$, assigns domains to clusters, so that the parameter controlling the vertical scale in covariance term, $k = 1$, for domain $i$ is $\theta^{\ast}_{1,s_{i},1}$.  The columns of the matrix,
\begin{align*}
    \mathbf{\Theta}^{\ast} = \kbordermatrix{\bm{\theta}^{\ast}& m~=~1&2&3&\\
    \theta^{\ast}_{1,m,1} & 2.61 & 0.38 & 0.91 \\
    \theta^{\ast}_{1,m,2} & 3.00 & 3.53 & 1.56 \\
    \theta^{\ast}_{2,m,1} & 1.04 & 2.26 & 0.84 \\
    \theta^{\ast}_{2,m,2} & 0.22 & 0.15 & 0.71
    }
\end{align*}
specify the set of $4$ locations generated for each of the $M = 3$ clusters that we used to construct $\{\mathbf{f}_{i}\}$.  Examining the length scale location parameters, $\theta^{\ast}_{\cdot,\cdot,2}$, in this table reveals that the surfaces in each cluster are formulated from the sum of a long length-scale or smooth term and a short length-scale or rough term.  So each surface is drawn from a mixture of two scales.  The long length scale locations were generated from, $\theta^{\ast}_{\cdot,\cdot,2} \sim \mathcal{G}a\left(3,2\right)$, while the short length-scale locations were drawn from, $\theta^{\ast}_{\cdot,\cdot,2} \sim \mathcal{G}a\left(2,5\right)$.  The vertical scale parameters for both terms were drawn from, $\theta^{\ast}_{\cdot,\cdot,1} \sim \mathcal{G}a\left(3,3\right)$.

The resulting observed surface for domain $i$, $\mathbf{y}_{i} \sim \mathcal{N}_{T}\left(\mathbf{f}_{i},\tau_{\epsilon}^{-1}\mathbb{I}_{T}\right)$, where the global noise precision parameter, $\tau_{\epsilon}$, is set to produce a $20\%$ noise-to-signal ratio (based on the average variance of the $\{\mathbf{f}_{i}\}$).

\begin{figure}[!h]
\begin{center}
\includegraphics[width=5.5in,height=2.8in]{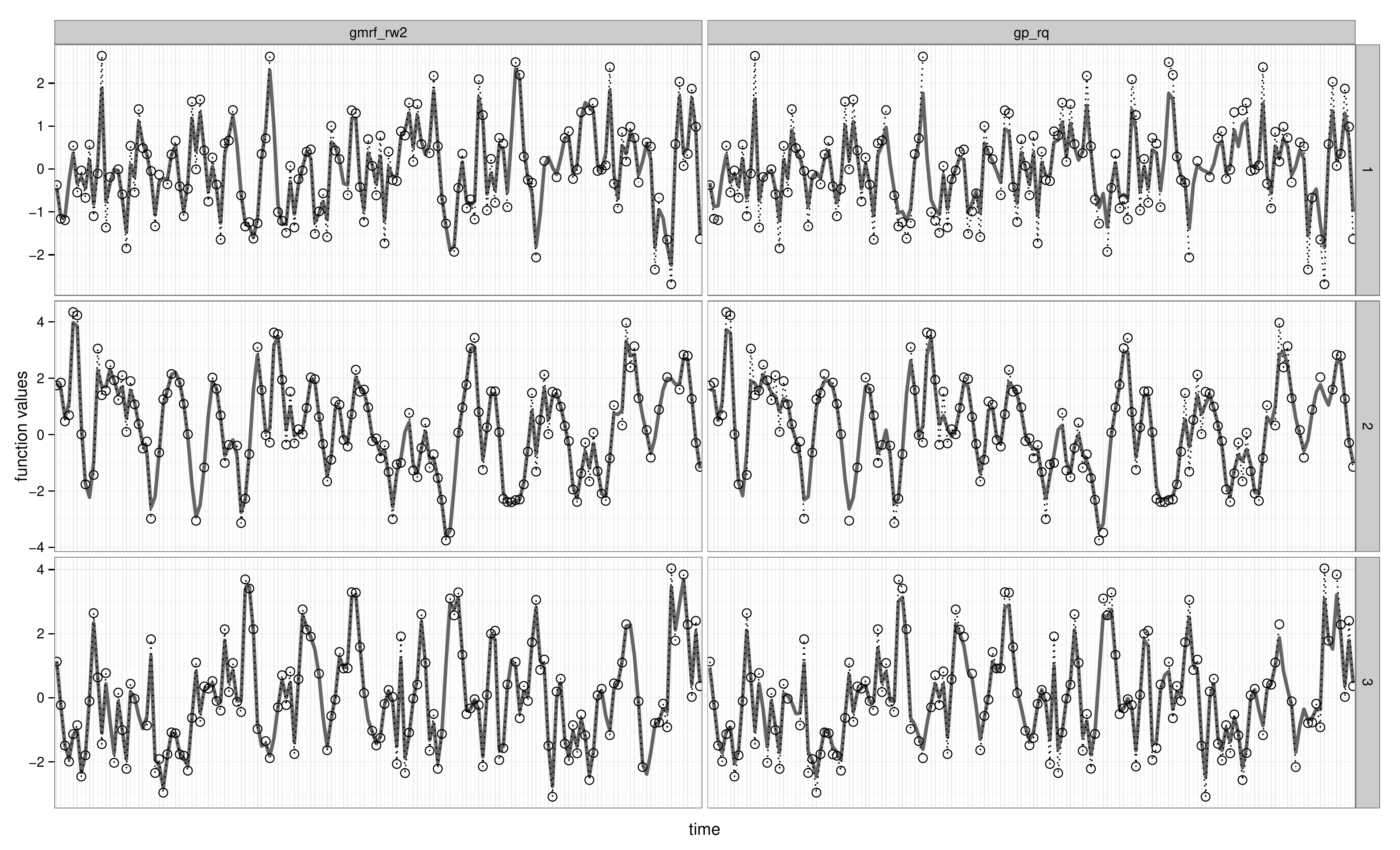}
\caption{Posterior mean values for estimated functions, $\{\mathbf{f}_{i}\}$, from the iGMRF
model, in the left-hand column, and under the GP, in the right-hand column under a simulated data set generated from a GP mixture of two squared exponential covariance terms.  The rows
represent each of the $M = 3$ clusters and each plot includes a randomly-selected domain.}
\label{2sefit}
\end{center}
\end{figure}

The first of our two primary inferential goals is to uncover the de-noised latent functions, $\{\mathbf{f}_{i}\}$, so we want to avoid over-fitting the observed noisy surfaces.  We perform estimation after randomly setting $10\%$ of the $NT$ observations to missing as presented to the estimation models to use them as a test set.  The GP and iGMRF models will estimate function values, $\{\hat{f}_{ij}\}$, for each of these missing values and we compute a mean-squared prediction error (MSPE) based on the squared difference between the estimated and true function values for the test set.  We then normalize the MSPE by the variance of the test set latent functions to produce a normalized MSPE.  Figure~\ref{2sefit} compares fitted results (solid lines) to data values (circles) for the iGMRF $RW2(\kappa)$ model, in the left column, with the GP formulation under a single rational quadratic covariance formula, in the right column.  The rows represent each of the $M = 3$ clusters and the plot in each cell is for a randomly-selected domain from the represented model-cluster combination.  We see that the GP model tends to attenuate the peaks to a greater degree than does the iGMRF.  The normalized MSPE values are $0.39$ for the GP and $0.54$ for the iGMRF. The GP fits notably better precisely because it avoids over-fitting.

Our second inferential goal is to utilize the clustering to provide context about how the functions compare among the observed domains (e.g. geographic or industry labels).  Both the GP and iGMRF models employ a DP mixture formulation under which a clustering is generated on each posterior sampling iteration that assigns the covariance or precision parameters, respectively, to clusters.  Both the number of clusters, $M$, and the allocations of domains to those clusters, may change on each posterior sampling iteration, so that these samples, taken together, formulate draws from the posterior distribution over the space of clusterings (or partitions).  The relative concentration of this distribution may be assessed by examining the degree of similarity in number of clusters and assignment of the domains to those clusters among the posterior draws.

We select \emph{one} clustering of domains from among the MCMC draws by using the least-squares algorithm of \citet{dahl:2006} that builds an $N \times N$ matrix of pairwise clustering probabilities to summarize the posterior draws and selects that clustering which is ``closest" to this matrix under a squared Euclidean distance metric.  Once we have selected a clustering, we build an $N\times N$ pairwise matrix, where cell $(i,j)$ is filled with a $1$ if states $i$ and $j$ are clustered together; otherwise, it is filled with a $0$.  We also build this pairwise clustering matrix for the true clustering used to generate the observed (synthetic) values.  We then compare the true and model-based pairwise matrices and compute a percent mis-clustering based on the sum of the cell values that don't agree.

\begin{figure}[!h]
\begin{center}
\includegraphics[width=3.5in,height=4.0in]{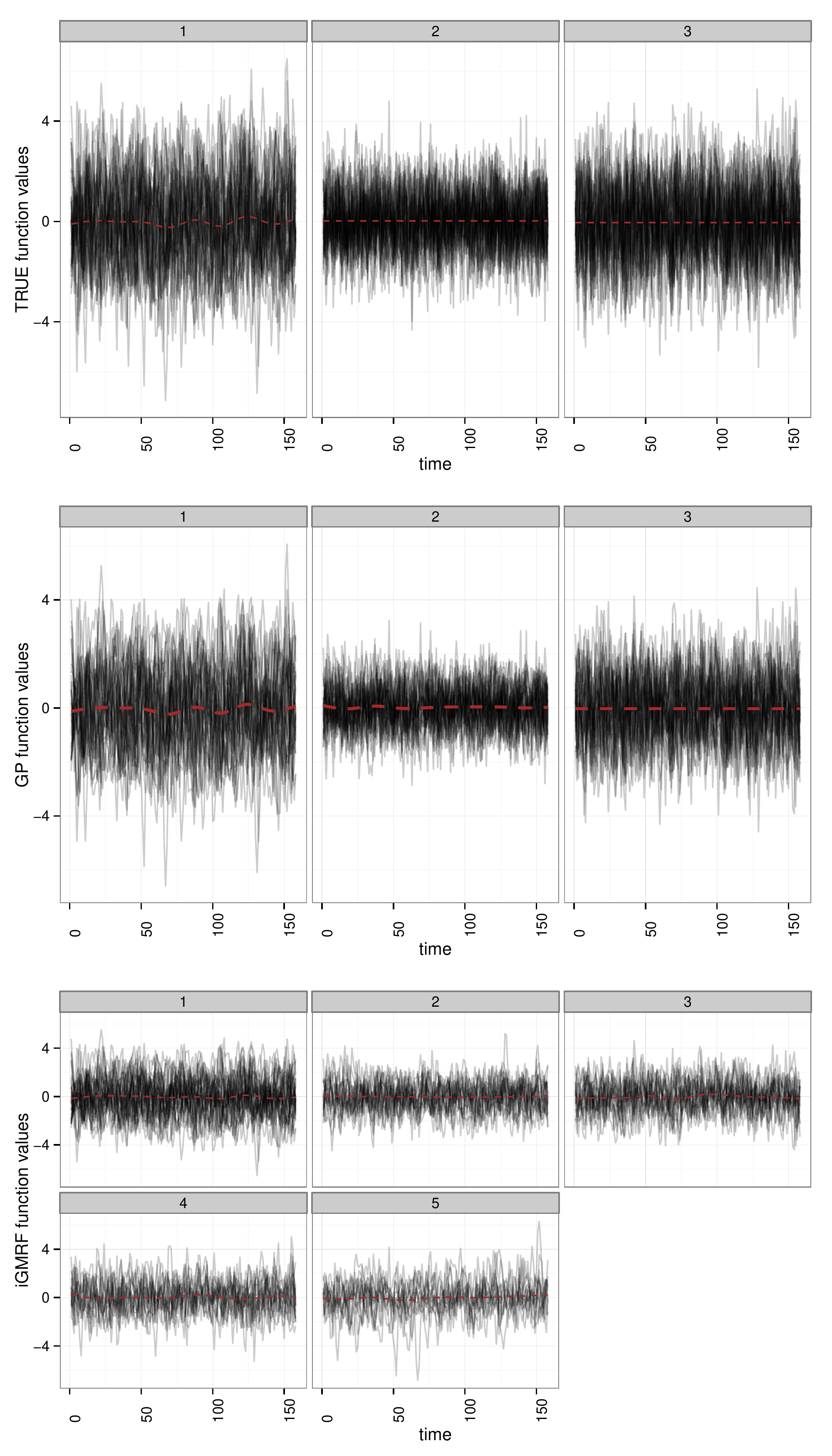}
\caption{Cluster sub-groupings of the posterior means for estimated functions, $\{\mathbf{f}\}$. Each panel plots the subset of functions assigned to a cluster under a simulated data set generated from a GP mixture of two squared exponential covariance terms.  A loess smoother is added in each panel with a red dotted line to highlight patterns among the assigned functions.  Each row represents the clustering or partition under a model.  The first row presents the true functions and clustering, while the second row presents the estimated clustering from GP model and the last two rows, the GMRF model.}
\label{2cluster}
\end{center}
\end{figure}

Figure~\ref{2cluster} presents a visual schematic of the true clustering for our synthetic data and the estimated clustering structure under each of the two models.  Each plot panel (in each of three groupings of panels) collects the posterior mean values for estimated functions assigned to the cluster for that panel.  Each grouping of panels represents a clustering under a different model.  The top row presents the true clustering under $3$ clusters used to generate the observed $\{\mathbf{y}_{i}\}$, while the second row presents the estimated clustering for the GP model (selected under the least squares algorithm) and the last two rows present the clustering for the iGMRF model.  We see that the GP model well reproduces the generating cluster with a mis-clustering error of only $2\%$.  The iGMRF model discovers the $3$ clusters, but also creates echoes of the third cluster, producing a total of $5$ clusters and a resulting mis-clustering error of $20\%$.

\begin{figure}[!h]
\begin{center}
\includegraphics[width=5.5in,height=2.8in]{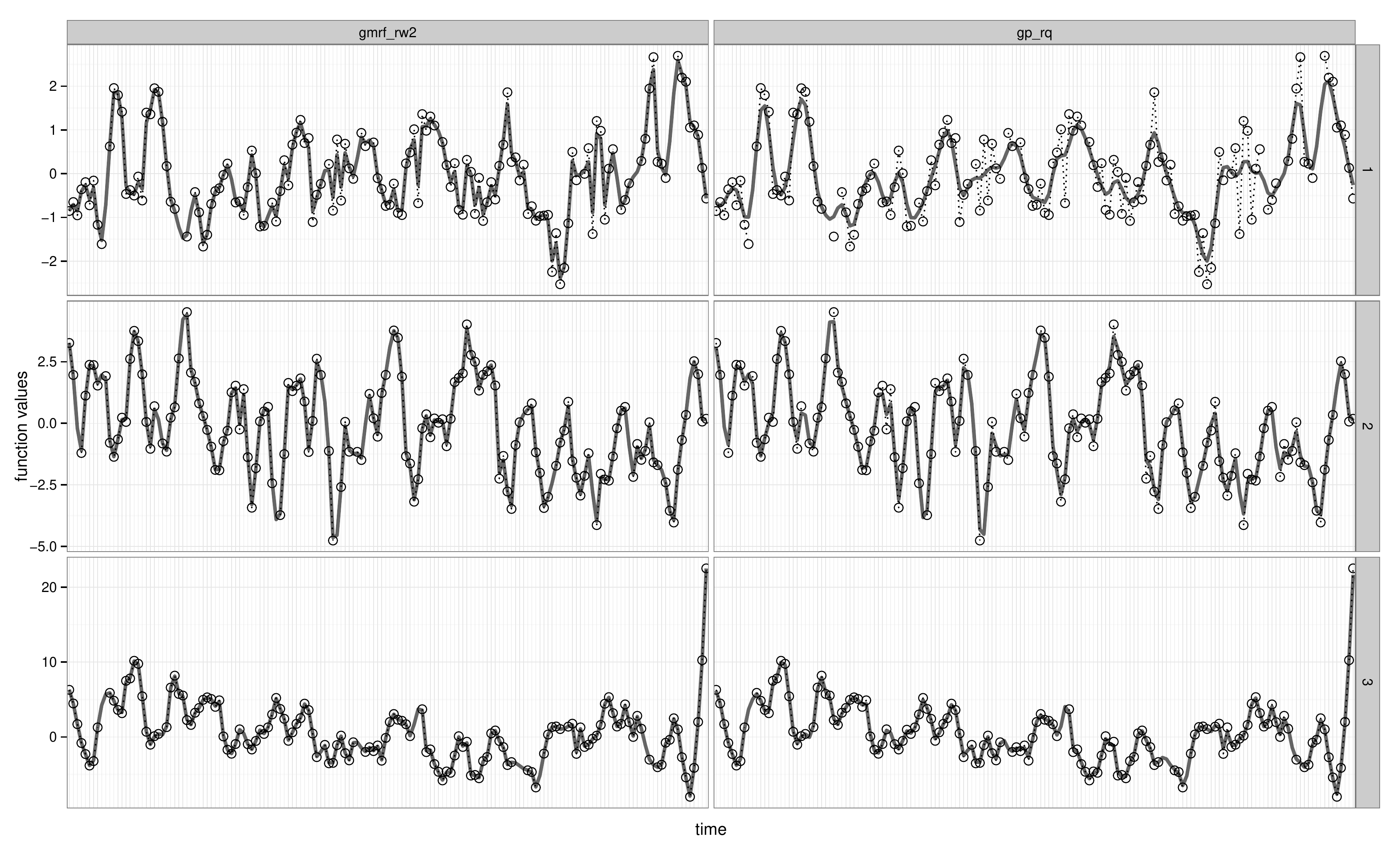}
\caption{Posterior mean values for estimated functions, $\{\mathbf{f}_{i}\}$, from the iGMRF
model, in the left-hand column, and under the GP, in the right-hand column under a simulated data set generated from a proper GMRF of order $2$ with strength-of-correlation, $\rho = 0.95$.  The rows represent each of the $M = 3$ clusters with a randomly-selected domain.}
\label{gmrffit}
\end{center}
\end{figure}

\begin{figure}[!h]
\begin{center}
\includegraphics[width=3.5in,height=4.0in]{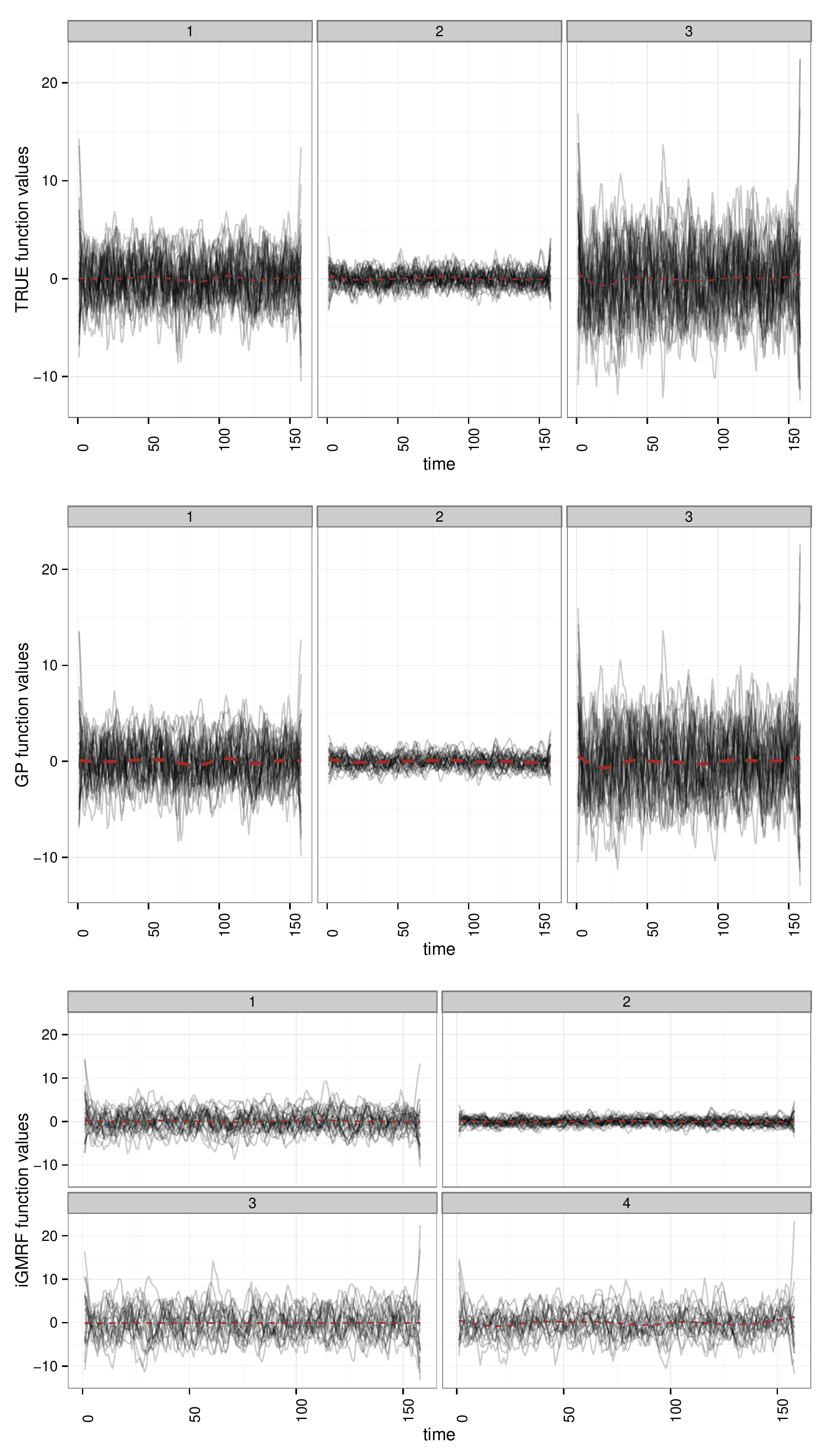}
\caption{Cluster sub-groupings of the posterior means for estimated functions, $\{\mathbf{f}\}$.  The estimation was performed on a simulated data drawn from a proper GMRF of order $2$ with strength-of-correlation, $\rho = 0.95$.  Each row represents the clustering or partition under an estimation model.  The first row presents the true functions and clustering, while the second row presents the GP model and third, the GMRF model.}
\label{gmrfcluster}
\end{center}
\end{figure}

Perhaps we expect the GP model with a rational quadratic covariance formulation to perform relatively better than the iGMRF model since the rational quadratic is designed for estimation under varying scales.  So our next simulated procedure generates latent functions from a model closer to the iGMRF.  Recall that that the $T\times T$ precision matrix of GMRF may be expressed as, $\mathbf{R}_{i} = \kappa^{\ast}_{s_{i}}\left(\mathbf{D}-\rho\mathbf{\Omega}\right)$, where we set the strength of correlation parameter, $\rho = 1$ (which produces a rank-deficient precision matrix), to produce our iGMRF formulation.  If we set $0 < \rho < 1$, we recover a full rank precision matrix, but at the expense of shrinking the conditional mean values to $0$.  We draw,
\begin{equation*}
\mathbf{f}_{i} \sim \mathcal{N}_{T}\left(\mathbf{0},\kappa^{\ast}_{s_{i}}\left(\mathbf{D}-0.95\mathbf{\Omega}\right)\right)
\end{equation*}
to construct our next synthetic dataset, under a second order specification for $\{\mathbf{D},\mathbf{\Omega}\}$, as described in Section~\ref{gmrfmod} using the same $N,T,M$ as for the first simulation.  Locations, $\{\kappa^{\ast}_{m}\}_{m=1,\ldots,3}$, are generated from a $\mathcal{G}a\left(1,1\right)$. As before, we set $\tau_{\epsilon}$ to produce a $20\%$ noise-to-signal ratio.  The resulting functions will be similar to those that would be generated under an iGMRF.  Figure~\ref{gmrffit} demonstrates that the GP formulation (under the rational quadratic covariance formulation) tends to produce a smoother fit, making it easier to separate signal from noise.  The fit performances are very similar between the GP and iGMRF models, with normalized MSPE values of $0.157$ and $0.159$, respectively.
Even though both models deliver similar fit accuracy, Figure~\ref{gmrfcluster} reveals that the GP formulation continues to do a better job of discovering the true clustering.  Again, both models discover the framework of the true clustering, though the iGMRF model tends to echo one of the clusters.  The mis-clustering rates here are $0\%$ and $12\%$ for the GP and iGMRF models, respectively, indicating that they both do a good job, but the GP model does well even when the underlying true functions are generated from a different process than a GP.  The GP employs $3$ parameters in it's covariance formula, giving it the flexibility to detect both the large vertical scale differences that differentiate the clusters and to model length scale variations within a function.  The iGMRF does better here than under the first simulation precisely because the clusters are primarily differentiated by (vertical) scale.

We conclude our simulation study by estimating the GP formulation of Equation~\ref{globalgp}, where we \emph{exclude} the modeling of clusters among $\{\mathbf{f}_{i}\}$.  We seek to examine the out-of-sample fit performance when we ignore the dependence of the time-indexed collection of functions among the domains.   We compare MSPE fit performance between including and excluding the clustering prior on the second simulated dataset.   The resulting normalized MSPE of $0.3$ under the model that excludes the clustering prior represents a substantial deterioration of fit from results under both the GP and GMRF constructions of Equations~\ref{dpgp} and \ref{dpgmrf}, which model the dependence among the domains.  So ignoring the dependence among domains (e.g. states) may reduce the accuracy of estimating the latent, de-noised functions, $\{\mathbf{f}_{i}\}$.

While the GP model demonstrates superior robustness, it comes at a computational cost.  Under our implementation of tempered transitions in a temporary space for sampling GP locations, we roughly achieve the same information (effective sample size) in $10000$ iterations of the GP model as we do in $25000$ iterations of the iGMRF, though former consumes $28900$ CPU-seconds and the later only $1226$ CPU-seconds on a single core of an Intel-i7-3450-powered laptop where the estimation routines are both written in C++.  The successively coarser covariance approximations for likelihood evaluations in the temporary space were computed with $(100,60)$ of the $T=158$ time points.  It is likely we could better if we were to optimize these choices (from the standpoint of computation time per effective sample size), though the large difference will remain.

\section{CPS Employment Application} \label{application}
We return to our motivating application for which we focus on reducing the noise-induced volatility in the $T = 158$ month series of employment totals for $N = 51$ states (including the District of Columbia) reported from the CPS.  Each state's observed employment series is standardized to remove the overall magnitude in order to facilitate comparisons across states based on similarities in the shapes and patterns expressed. States seek context for changes in employment totals as they filter down to the county-level estimates, so we will perform inference on the distribution over clusterings to understand which states express similarities in their employment patterns.

\begin{figure}[!h]
\begin{center}
\includegraphics[width=5.0in,height=4.0in]{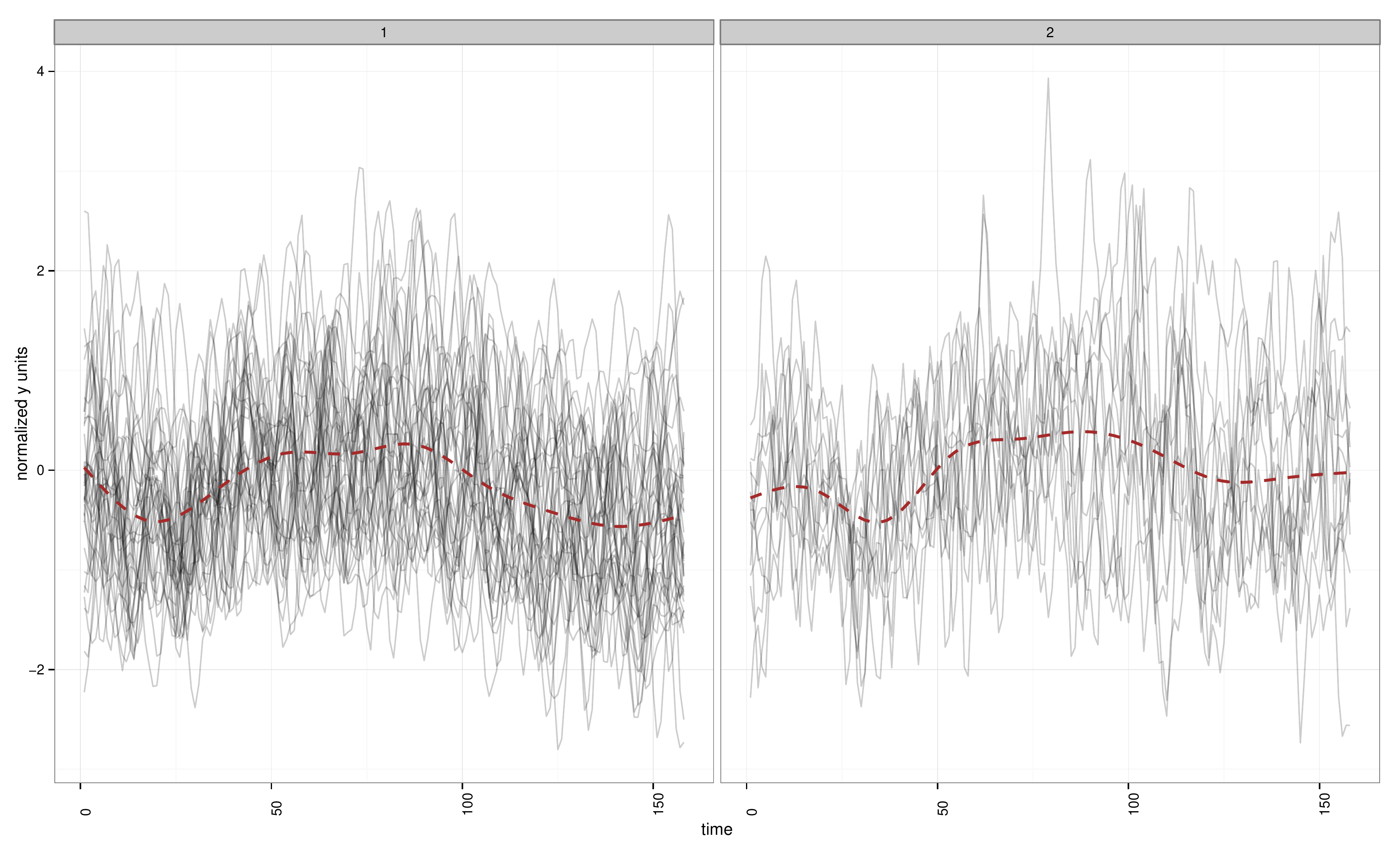}
\caption{Posterior mean estimated functions, $\{\mathbf{f}\}$, grouped into clusters.  Each panel plots the set of functions assigned to a cluster for the CPS state-level dataset under the selected least squares clustering.  A loess smoother is added in each panel with a red dotted line to highlight a pattern among the assigned functions.}
\label{cescluster}
\end{center}
\end{figure}

\begin{figure}[!Ht]
\begin{center}
\includegraphics[width=6.8in,height=4.5in]{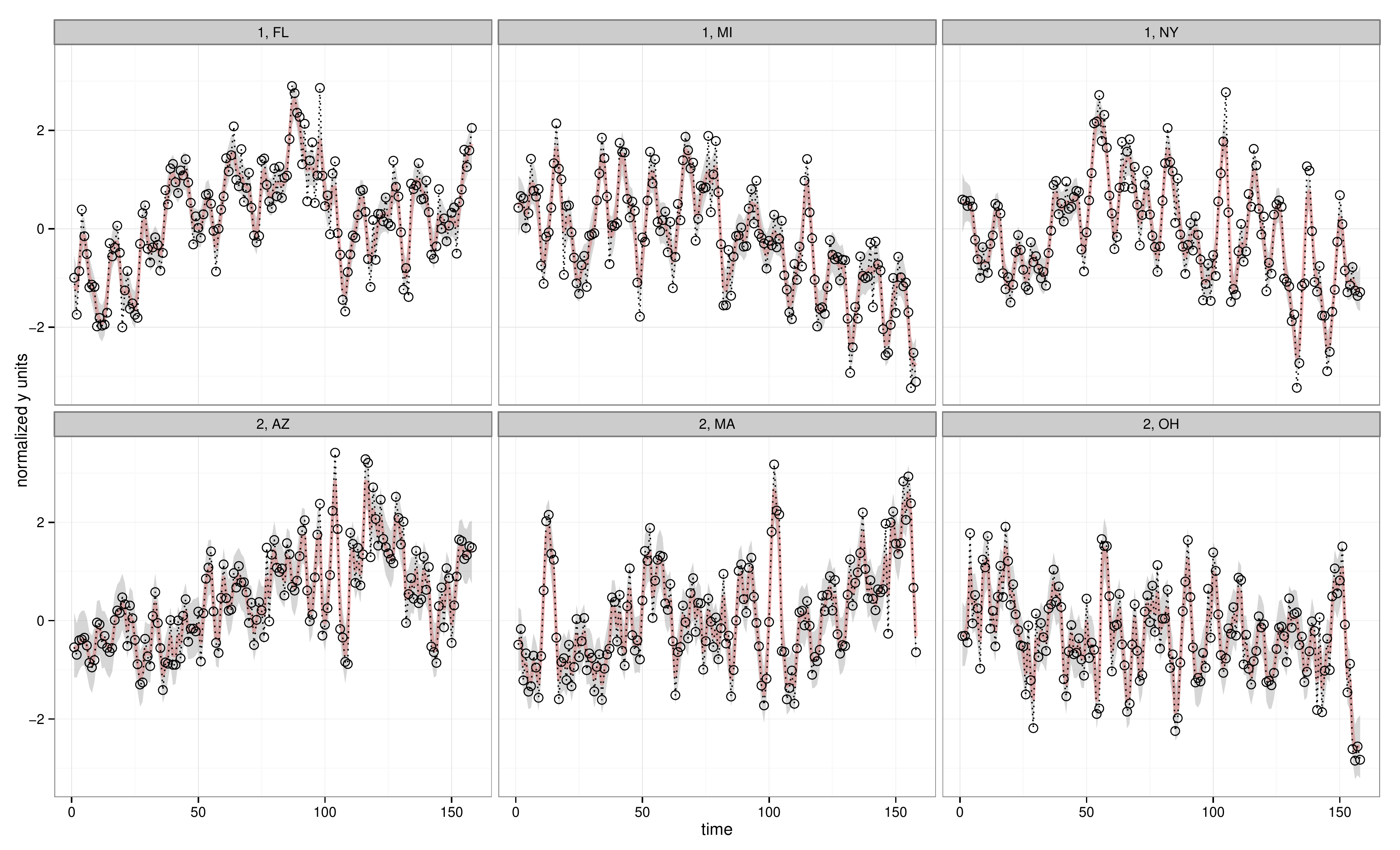}
\caption{Posterior mean values for estimated functions, $\{\mathbf{f}\}$, from the mixtures of GPs model, represented by the pink lines, with $95\%$ credible intervals displayed as gray shading compared to the noisy series, $\{\mathbf{y}\}$, represented by the black hollow circles, connected with black dashed lines. Each panel represents a series for a single state.  The state plots in the top row are assigned to the left-hand cluster of Figure~\ref{cescluster}, while the bottom row plots are
from the right-hand cluster.}
\label{cesmultifit}
\end{center}
\end{figure}

We fit the GP mixture model of Equation~\ref{dpgp} to the CPS data.  Figure~\ref{cescluster} presents the allocation of estimated latent functions, $\{\mathbf{f}_{i}\}$, into assigned clusters of the clustering selected under the least squares criterion of \citet{dahl:2006}.  We recall that the least squares algorithm selects a single clustering from among the set of posterior samples for the clustering.  Each panel of the figure displays the estimated functions for those states assigned to each of $2$ clusters estimated by the model. Examining the pattern in the collection of functions assigned to each cluster reveals that the clusters are primarily differentiated by the employment sensitivities expressed by member states to the ``great recession" that began in approximately $2008$ (at month $96$).  The latent functions for those states included in the left-hand panel express a notably longer lasting trough than those states included in the right-hand panel (where panels index cluster memberships).  The initial rate of employment drop appears similar in both clusters, but the decline continues longer and recovers more slowly among states assigned to the left-hand cluster. We conducted additional runs that varied the shape hyperparameter of the $DP$ concentration prior on $\alpha \sim \mathcal{G}a\left(a_{4},1\right)$, with $a_{4} \in \{1 - 4\}$ to induce successively higher prior number of clusters, but the estimated models all performed very similarly and the $M = 2$ clustering was selected.  Estimation under the iGMRF mixture model of Equation~\ref{dpgmrf} produces the same selected clustering.
The left-hand cluster includes a number of states that experienced so-called bubbles in property values, such as California, Florida, Hawaii, and Nevada, as well as states sensitive to both financial services and auto industries, such as New York and Michigan.  Table~\ref{tab:cluster} lists the $38$ states assigned to the left-hand cluster of Figure~\ref{cescluster} and the $13$ states assigned to the right-hand cluster.
\begin{table}[h]
\begin{center}
\begin{tabularx}{\linewidth}{lX}
{\bf Cluster} & {\bf Member States}\\
\hline
{\bf Left} & AK, AL, AR, CA, CO, CT, DC, FL, GA, HI, IA, IL, IN, KS, KY, LA, MD, MI, MN, MO, MS, ND, NE, NV, NY, OK, OR, PA, RI, TX, UT, VA, VT, WV, WY\\
{\bf Right} & AZ, DE, ID, MA, MT, NC, NJ, OH, SC, SD, TN, WA, WI
\end{tabularx}
\caption{State Memberships in $2$ Estimated Clusters \label{tab:cluster}}
\end{center}
\end{table}
Each panel in Figure~\ref{cesmultifit} presents the fitted latent functions from the GP model of Equation~\ref{dpgp} with a pink line and the associated credible intervals shaded in gray, compared to the actual observed data values.  The states represented in the top row of panels are assigned to the left-hand cluster of Figure~\ref{cescluster}, while the states presented in the bottom row are assigned to the right-hand cluster.  Examining the states assigned to the left-hand cluster, both New York and Michigan experienced a relatively more rapid descent in employment, despite the large volatility in New York, while Florida appears to have experienced a large, nearly sudden, drop followed by a stabilization and recovery.  Contrastingly, the Arizona and Massachusetts appear to have rather rapidly bounced up from the initial drop, while Ohio experienced a less dramatic fall in employment.  The cluster memberships provide state administrators context on the common economic drivers for their employment trends through the great recession.
\begin{figure}[!Ht]
\begin{center}
\includegraphics[width=6.8in,height=4.5in]{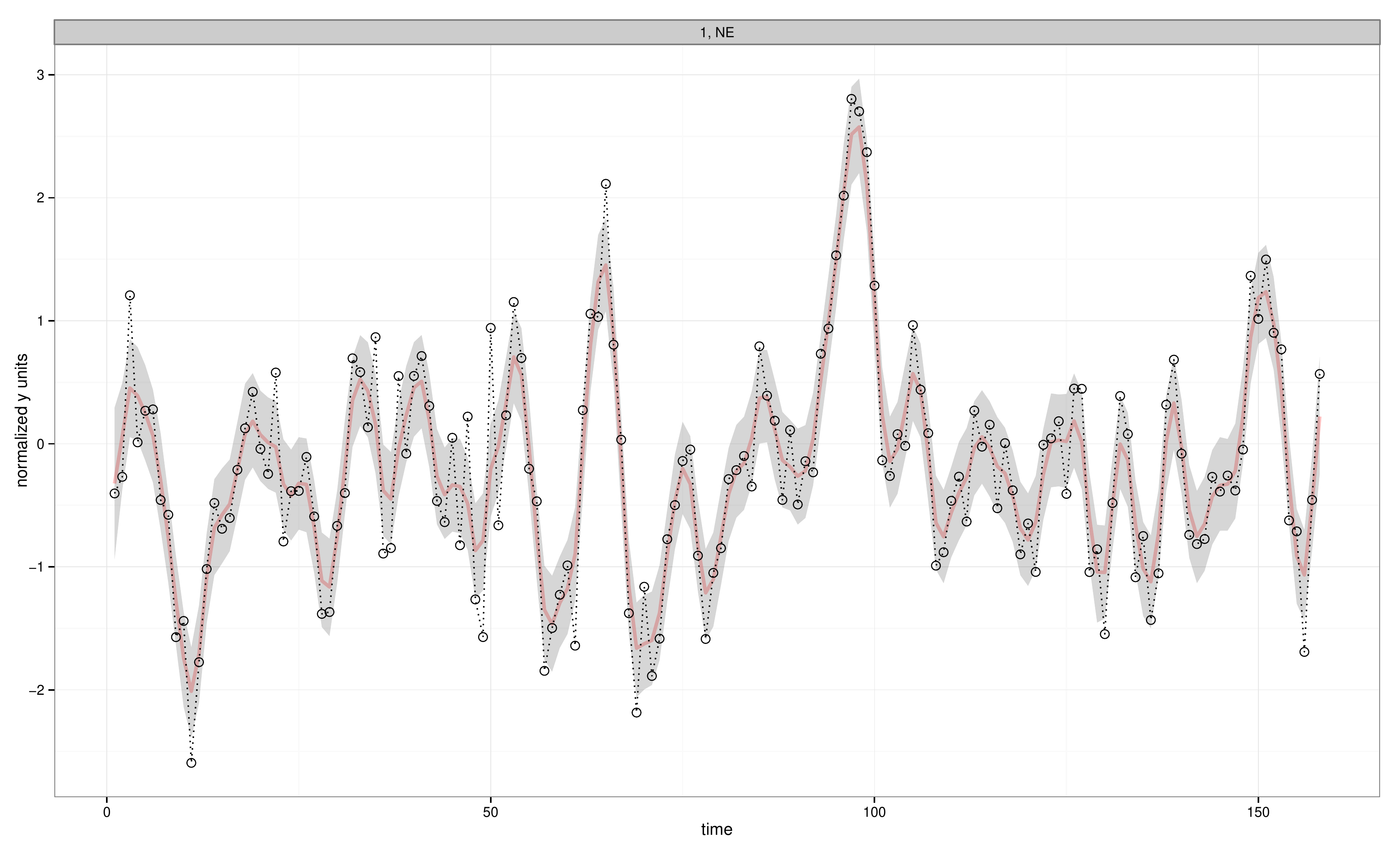}
\caption{Posterior mean estimated function for the state of Nebraska from the mixtures of GPs model, represented by the pink line, with $95\%$ credible intervals displayed as gray shading compared to the noisy series, $\{\mathbf{y}\}$, represented by the black hollow circles, connected with black dashed lines.}
\label{ces:ne}
\end{center}
\end{figure}

Lastly, the solid, pink line in Figure~\ref{ces:ne} displays the de-noised estimated function for the state of Nebraska compared to the noisy, CPS estimates (displayed in the hollow circles connected with a dashed line) to demonstrate the reduction of noise accomplished under our mixtures of GPs model.

\section{Discussion} \label{discussion}
Our task was to estimate latent, state-indexed functions that separate away noise-induced volatility present in our observed CPS data to improve the efficiency and interpretability of resulting county-level employment statistics constructed from the state estimates.  We developed nonparametric mixture formulations that simultaneously estimate the latent, state-indexed functions and allow the data to discover a general dependence structure among them.  Our simulation study results demonstrated that failing to account for a dependence structure among states in the estimation model deteriorates the ability to uncover the true latent functions.

Our DP mixture of GPs or iGMRFs, outlined in Equations~\ref{dpgp} and \ref{dpgmrf}, employ an unsupervised approach for discovering dependence among the state functions based on similarities in the time-indexed patterns expressed in the data.  They perform well on our CPS employment count application, uncovering a differentiation among states based on their employment sensitivities to the great recession.  The simulation study revealed a greater robustness, both in estimation of accuracy of the latent functions and their clustering properties, of the GP mixture model as compared to the iGMRF mixture, due to the regulated smoothness properties of the rational quadratic covariance formulation and its employment of more parameters than under the iGMRF that control for the trend, scale and frequency properties of the estimated latent functions.  The iGMRF computes much faster, however, so it may still be useful, particularly in the case where the clusters are differentiated based primarily on vertical magnitude of the functions.  The computational intensity of the GP mixture is mitigated with our adaptation of the faster-computing MCMC algorithm of \citet{wang:2013} (which uses tempered transition steps in a temporary, lower-dimensional space) from a single GP construction to our nonparametric mixtures of GPs model.

Our methodology may be extended by employing domain level predictors to help determine the distribution over clusterings with an approach such as the probit stick-breaking process of \citet{Rodr:Duns:nonp:2011}.  This approach would replace the single unknown measure, $G$, imposed on the parameters of the covariance or precision matrix with an uncountable collection of unknown measures, $\{G_{\mathbf{x}}\}_{\mathbf{x} \in \mathbb{R}^{d}}$, of such measures based on the values for $d$ predictors.  The collection of measures is induced by indexing the weights of the stick breaking construction of Equation~\ref{stick} with $\mathbf{x}$.  This approach may be expected to more precisely estimate the dependence structure in the case of a \emph{large} number of domains (such as counties, rather than states) on the order of $1000$ or more.

\section{Acknowlegdements}
The authors wish to thank our colleagues at the Bureau of Labor Statistics whose focus on continuous improvement led them to sponsor this project.  We thank the following important contributors:
\begin{enumerate}
\item Sean B. Wilson, Senior Economist, who formulated the project and sponsored our work.
\item Garrett T. Schmitt, Senior Economist, who helped us think through alternative approaches.
\item Bradley A. Jensen, Senior Economist, who provided us multiple data slices that allowed us to perform our estimation.
\end{enumerate}
\bibliographystyle{elsarticle-harv}
\bibliography{mv_refs_jul2014}

\newpage
\appendix
\section{Posterior Sampling Algorithms}\label{AppMain}
\subsection{Gaussian Process Sequential Scan}
We sample the parameters of the mixtures of GPs model, $\bm{\lambda} = \left(\mathbf{\Theta}^{\ast},\mathbf{s},\alpha,\tau_{\epsilon}\right)$, in a sequential scan from the full conditionals after marginalizing over the latent, de-noised functions, $\{\mathbf{f}_{i}\}$, which is easy to do and leads to slightly better mixing.  We must, however, co-sample the $\{\mathbf{f}_{i}\}$ in the case of intermittent missingness-at-random in $\{\mathbf{y}_{i}\}$ in order to allow the sampling of missing values from their posterior predictive distribution.   We first outline our sequential scan in the case where we marginalize out the latent functions and conclude this section by highlighting changes required in the case we co-sample them.  The mixtures of GPs model specification is highly non-conjugate.  We carefully configure blocks of parameters to permit application of posterior sampling approaches designed to produce robust chain mixing under non-conjugate specifications.
\begin{enumerate}
\item Sample cluster locations, $\{\theta^{\ast}_{pm}\}_{p = 1,\ldots,P;~m = 1,\ldots,M}$, (where $p$ denotes parameter type (e.g. $\left(\theta^{\ast}_{1},\theta^{\ast}_{2},\theta^{\ast}_{3}\right)$) and $m = 1,\ldots,M$ denotes the cluster) and model error precision, $\tau_{\epsilon}$:
  We sample the posterior distribution for locations in by-cluster groups, $\{\theta^{\ast}_{pm}\}_{p = 1,\ldots,P}$, from the subset of observations (states) assigned to that cluster because $\theta^{\ast}_{pm} \independent \theta^{\ast}_{pm^{'}}$ for $m^{'} \neq m$, \emph{a posteriori}, in a Metropolis-Hastings scheme using the following log-posterior kernel,
    \begin{eqnarray}
       &&\mbox{log}~\pi\left(\theta^{\ast}_{pm}|\bm{\theta}^{\ast}_{-pm},\mathbf{s},
        \tau_{\epsilon},
        \{\mathbf{y}_{i}:s_{i} = m\}\right) \nonumber\\
         &&\propto -\frac{1}{2}n_{m}\mbox{log}\left(|\mathbf{C}^{\tau}\left(\bm{\theta}_{m}^{\ast}\right)|\right)
        -\frac{1}{2}\sum_{i \in s_{m}} \mathbf{y}_{i}^{'}\mathbf{C}^{\tau}\left(\bm{\theta}_{m}^{\ast}\right)\mathbf{y}_{i} + (a-1)\log(\theta^{\ast}_{pm}) - b\theta^{\ast}_{pm},\label{posttheta}
    \end{eqnarray}
    where $\mathbf{C}^{\tau}\left(\bm{\theta}_{m}^{\ast}\right) = \mathbf{C}\left(\bm{\theta}_{m}^{\ast}\right) + \left(1/\tau_{\epsilon}\right)\mathbb{I}_{T}$,  $s_{m}$ collects the states assigned to cluster $m$, $n_{m}$ denotes the number of states assigned to cluster $m$, and $(a,b)$ are shape and rate hyperparameters of a gamma prior, respectively.  This posterior representation is a relatively straightforward Gaussian kernel of a non-conjugate probability model.

    We adapt a Metropolis-Hastings algorithm of \cite{wang:2013} designed to speed computation for sampling \emph{each} $\theta^{\ast}_{pm}$ (and also, $\tau_{\epsilon}$) by introducing a lower-dimensional temporary space where the likelihood (e.g. the $T \times T$, Gaussian process covariance matrix, $\mathbf{C}$) is approximated using a subset of the $T$ time-points.  Starting with the previously sampled value, $\hat{x}_{0}$, (for $\theta^{\ast}_{pm}$ or $\tau_{\epsilon}$) from the full-dimensional or exact space, the sampling algorithm builds a sequence of proposed transitions by first ``stepping up" in the temporary space using increasingly coarse, ``tempered" transitions, $\left(\hat{Z}_{1},\hat{Z}_{2},\ldots,\hat{Z}_{n}\right)$, that generate computationally-fast approximations for $\mathbf{C}$.  These approximations use fewer observations in each step; for example, we apply $n = 2$ transitions in the lower-dimensional space and use $100$ of the $T = 158$ time points to formulate $\hat{Z}_{1}$ and then $60$ time points for $\hat{Z}_{2}$.  This sequence of coarser transitions is followed by transition steps that employ progressively finer distributions (in reverse order), $\left(\check{Z}_{2},\check{Z}_{1}\right)$  that "step down" to guide the chain back towards the full dimensional space until we conclude the sequence by proposing $\check{x}_{0}$ from $\check{Z}_{1}$ to be evaluated in the full-dimensional space.  We use univariate slice sampling of \citet{Neal00slicesampling} to accomplish these (reversible) transitions in the lower-dimensional space.  The proposal (reversible sequence of steps) is then accepted with (for $n = 2$ tempered transitions) probability of move formulation,
    \begin{equation*}
    \min\left(\frac{\pi_{1}(\hat{x}_{0})\pi_{2}(\hat{x}_{1})\pi_{1}(\check{x}_{1})}{\pi_{1}(\hat{x}_{1})
    \pi_{2}(\check{x}_{1})\pi_{1}(\check{x}_{0})}\frac{\pi(\check{x}_{0})}{\pi(\hat{x}_{0})}\right),
    \end{equation*}
    where ``$x$" denotes the proposals, and ``$\pi$" posterior kernel evaluations, for each $\theta^{\ast}_{pm}$ with subscript $0$ pertaining to the exact space and $(1,2)$ to the successively coarser transition distributions in the temporary space in the sequential order of application.  If our lower dimensional approximations are relatively good, this approach will speed chain convergence by producing draws of lower autocorrelation since each proposal includes a sequence of moves generated in the temporary space.  Since the moves in the temporary space are executed with fast approximations, \cite{wang:2013} show that this algorithm has the potential to substantially reduce computation time, as compared to the usual Metropolis-Hastings algorithm, for drawing an equivalent effective sample size.  The probability of move formulation evaluates the proposals on the full space of size $T$, however, so that the resulting sampled draws are from the \emph{exact} posterior distribution, rather than from a sparse approximation.

    Our adaptation configures \cite{wang:2013} to apply to clusters of Gaussian processes (rather than to just a single GP) by exploiting the between cluster posterior independence of the $\{\theta^{\ast}_{pm}\}$.
\item Sample cluster assignments, $\mathbf{s} = \left(s_{1},\ldots,s_{N}\right)$:
    We marginalize over $G$ in Equation~\ref{dpgp}, that results in the P\'{o}lya urn representation of \citet{blackwell:1973}, to sample $\mathbf{s}$ from its full conditionals,
    \begin{equation}\label{posts}
        \pi\left(s_{i}=s| \mathbf{s}_{-i},\mathbf{\Theta}^{\ast},\alpha,
        \tau_{\epsilon},\mathbf{y}_{i}\right) \propto
        \begin{cases}
        \frac{n_{-i,s}}{n-1+\alpha}\mathcal{N}_{T}\left(\mathbf{y}_{i}|\mathbf{0},
        \mathbf{C}^{\tau}\left(\bm{\theta}^{\ast}_{s}\right)\right) & \text{if $1 \leq s  \leq M^{-}$}\\
        \frac{\alpha/c^{*}}{n-1+\alpha}\mathcal{N}_{T}\left(\mathbf{y}_{i}|\mathbf{0},
        \mathbf{C}^{\tau}\left(\bm{\theta}^{\ast}_{s}\right)\right) & \text{if $s = M^{-} + h$},
        \end{cases}
    \end{equation}
    where $n_{-i,s} = \sum_{i^{'} \neq i}\mathbf{1}(s(i^{'}) = s)$ is the number of states, excluding state $i$, assigned to cluster $s$, so that states are assigned to an existing cluster with probability proportional to its ``popularity".  The posterior assigns an state (through $s_{i}$) to a new cluster with probability proportional to $\alpha d_{0}$ under the mixture prior in the case of a conjugate formulation.  The conjugate specification requires the likelihood to be integrable in closed form with respect to the base distribution, $G_{0}$, to compute $d_{0} = \int \mathcal{N}\left(\mathbf{y}|\bm{\theta},\ldots\right)G_{0}(d\bm{\theta})$, which is not the case under a (nonconjugate) GP construction.  So we employ the auxiliary Gibbs sampler formulation of \citet{neal:2000} and sample $c^{\ast} \in \mathbb{N}$ locations from base distribution, $G_{0}$, \emph{ahead} of any assigned observations (e.g. not yet linked to any state), to define $h = M^{-} + c^{\ast}$ candidate clusters in an augmented space.  We then draw $s_{i}$ from this augmented space, where any location not assigned states (over a set of draws for $\mathbf{s}$) is dropped.  This procedure effectively performs a Monte Carlo integration of the likelihood with respect to the base distribution.  See \cite{savitsky:2010} for a detailed example of a DP implementation on a GP. The larger is the tuning parameter, $c^{\ast}$, the lower is the autocorrelation of the resulting posterior draws (though computation time increases because we are sampling with respect to more cluster locations).  Good mixing is typically achieved with $c^{\ast}$ set equal to $2$ or $3$.  We note that the number of clusters, $M$, may change in this step as clusters are added and deleted.
\item Sample DP concentration parameter, $\alpha$:
    We use the algorithm of \cite{escobar:1995} that formulates the posterior for $\alpha$ as a mixture of two Gamma distributions with the mixture component, $\eta$, drawn from a beta distribution. The algorithm is facilitated by the conditional independence of $\alpha$ from data, $\mathbf{Y}$, given cluster assignments, $\mathbf{s}$, (that implies number of clusters, $M$).
\item Draw $\mathbf{f}_{i}$ as post-processing step from its posterior predictive distribution:
Use the Gaussian joint distribution for $\left(\mathbf{f}_{i},\mathbf{y}_{i}\right)$ to formulate the conditional posterior predictive Gaussian density,
\begin{equation}
\mathbf{f}_{i}\vert \mathbf{y}_{i} \sim \mathcal{N}_{T}\left(\mathbf{m}_{i},V_{i}\right),
\end{equation}
where $\mathbf{m}_{i} = \mathbf{C}\left(\bm{\theta}^{\ast}_{s_{i}}\right)[\mathbf{C}^{\tau}\left(\bm{\theta}^{\ast}_{s_{i}}\right)]^{-1}
\mathbf{y}_{i}$ and $V_{i} = \mathbf{C}\left(\bm{\theta}^{\ast}_{s_{i}}\right) -   \mathbf{C}\left(\bm{\theta}^{\ast}_{s_{i}}\right)[\mathbf{C}^{\tau}\left(\bm{\theta}^{\ast}_{s_{i}}\right)]^{-1}
\mathbf{C}\left(\bm{\theta}^{\ast}_{s_{i}}\right)$.
\item In the case of missing data values, we co-sample $\{\mathbf{f}_{i}\}$ (rather than marginalizing over them) after the cluster locations in a Gibbs scan from,
    \begin{equation*}
    \pi\left(\mathbf{f}_{i}|\mathbf{y}_{i},\mathbf{\Theta}^{\ast},\tau_{\epsilon}\right) = \mathbf{N}_{T}\left(\bm{\phi}^{-1}_{i}\mathbf{e}_{i},\bm{\phi}^{-1}_{i}\right),
    \end{equation*}
    where $T\times 1,~\mathbf{e}_{i} = \mathbf{f}_{i}^{'}\tau_{\epsilon}\mathbf{y}_{i}$ and $\bm{\phi}_{i} = \tau_{\epsilon}\mathbb{I}_{T} + \mathbf{C}\left(\bm{\theta}^{\ast}_{s_{i}}\right)^{-1}$, where we cache the cholesky decompostion of $\mathbf{C}$ from sampling locations, $\mathbf{\Theta}^{\ast}$, for faster computation of $\mathbf{C}^{-1}$.  Under co-sampling of the $\{\mathbf{f}_{i}\}$, we draw the model error precision,$\tau_{\epsilon}$, in a Gibbs step (instead of in the Metropolis-Hastings scheme along with $\mathbf{\Theta}^{\ast}$) using,
    \begin{equation*}
    \pi\left(\tau_{\epsilon}|\mathbf{Y},\{\mathbf{f}_{i}\}\right) = \mathcal{G}\left(a_{1},b_{1}\right),
    \end{equation*}
    where shape hyperparameter, $a_{1} = 0.5TN + a$ and rate hyperparameter, $b_{1} = 0.5\mathop{\sum}_{i=1}^{N}\left(\mathbf{y}_{i}-\mathbf{f}_{i}\right)^{'}
    \left(\mathbf{y}_{i}-\mathbf{f}_{i}\right) + b$ and $(a,b)$ are the prior hyperparameter settings.
    Lastly, we replace $\mathbf{y}_{i}$ with $\mathbf{f}_{i}$ under co-sampling of $\mathbf{f}_{i}$ in Equations~\ref{posttheta} and \ref{posts} for sampling cluster locations, $\mathbf{\Theta}^{*}$, and cluster memberships, $\mathbf{s}$.
\end{enumerate}

\subsection{Intrinsic Gaussian Markov Random Field Gibbs Scan}
Posterior sampling from the mixtures of iGMRFs employs a Gibbs scan over $\bm{\lambda} = \left(\{\mathbf{f}_{i}\},\{\kappa^{\ast}_{m}\},\mathbf{s},\alpha,\tau_{\epsilon}\right)$, from a set of conjugate full conditionals in the following steps:
\begin{enumerate}
\item Sample latent functions, $\{f_{ij}\}_{i=1,\ldots,N;~j = 1,\ldots,T}$ in a Gibbs steps from the full conditionals,
    \begin{equation*}
    \pi\left(f_{ij}\vert \{f_{ij}\}_{-j},\tau_{\epsilon}, y_{ij}\right) =
    \mathcal{N}\left(\frac{e_{ij}}{\phi_{ij}},\phi_{ij}^{-1}\right),
    \end{equation*}
    where $e_{ij} = \tau_{\epsilon}y_{ij} + Q_{jj}\kappa^{\ast}_{s_{i}}\bar{f}_{ij}$,
    with $\bar{f}_{ij} = -\frac{1}{Q_{jj}}\mathop{\sum}_{k:k\sim j}Q_{jk}f_{ik}$ and $\phi_{ij} = \tau_{\epsilon} + Q_{jj}\kappa^{\ast}_{s_{i}}$.
\item Sample locations, $\{\kappa^{\ast}_{m}\}$, in a Gibbs step from,
    \begin{equation*}
    \pi\left(\kappa^{*}_{m}|\{f_{ij}\}\right) = \mathcal{G}a\left(a_{2},b_{2}\right),
    \end{equation*}
    with shape parameter, $a_{2} = \frac{1}{2} n_{m}(T-o_{Q}) + a$, where $o_{Q} = 2$ is the rank-deficiency of the precision matrix, $\mathbf{Q}$, indicating that that  $\mathbf{f}_{i}$ provides the equivalent of $T-o_{Q}$ degrees of freedom, rather than number of time points, $T$.  The rate parameter, $b_{2} = \frac{1}{2}\mathop{\sum}_{i:s_{i} = m}\left(\mathop{\sum}_{j=1}^{T}Q_{jj}[f_{ij} - \bar{f}_{ij}]^{2}\right) + b$, where the rate parameter is composed from the subset of latent functions, $\{\mathbf{f}_{i}\}_{i:s_{i} = m}$, for those domains assigned to cluster $m$.
\item Sample cluster assignments under a similar P\'{o}lya urn representation as for the GP, only here the mixture posterior is conjugate, so,
    \begin{equation}
        \pi\left(s_{i}=s| \mathbf{s}_{-i},\mathbf{\kappa}^{\ast},\alpha,
        \tau_{\epsilon},\mathbf{f}_{i}\right) \propto
        \begin{cases}
        \frac{n_{-i,s}}{n-1+\alpha}\mathop{\prod}_{j=1}^{T}\mathcal{N}\left(f_{ij}|\bar{f}_{ij},
        [\kappa^{\ast}_{s_{i}}Q_{jj}]^{-1}\right) & \text{if $1 \leq s  \leq M^{-}$}\\
        \frac{\alpha}{n-1+\alpha}d_{0,i} & \text{if $s = M^{-} + 1$},
        \end{cases}
    \end{equation}
    where $d_{0,i} = \int \mathcal{N}\left(y|\kappa,\ldots\right)G_{0}(d\kappa) = \frac{b^{a}\Gamma(a_{2})\left[\mathop{\prod}_{j=1}^{J}\mathcal{N}\left(0|0,Q_{jj}^{-1}\right)\right]}
    {\Gamma(a)b_{2,i}^{a_{2}}}$, where $a_{2}$ is as defined above and $b_{2,i}$ is term $i$ in the sum that composes $b_{2}$, defined above.
\end{enumerate}
Finally, we sample the DP concentration parameter, $\alpha$, and the model noise precision, $\tau_{\epsilon}$, in the same fashion as for the GP.  Unlike the mixtures of GPs, we specify this model in a conjugate formulation that allows for fast sampling. 

\label{lastpage}
\end{document}